\documentstyle[sprocl]{article}

\def\be{\begin{equation}}
\def\ee{\end{equation}}
\def\bea{\begin{eqnarray}}
\def\eea{\end{eqnarray}}

\begin{document}

\title{European accelerator-based neutrino projects}

\author{Mario Campanelli}

\address{DPNC University of Geneva, 24 Quai E.Ansermet, CH-1211 Geneva Switzerland\\E-mail: Mario.Campanelli@cern.ch}

\maketitle\abstracts{Future neutrino projects in Europe will follow two
distinct time lines. On the medium term, they will be dominated by the
CERN-Gran Sasso long-baseline project, with two experiments OPERA and
ICARUS, mainly concentrated on $\tau$ appearance. On the longer term,
several projects are under discussion. A new proton driver at CERN
that accelerates a 4 MW beam to 2.2 GeV of energy would open the
possibility of a low-energy super-beam, possibly sent to the French
laboratory under the Frejus. A new radioactive heavy ion facility could
produce a pure $\nu_e$ beam, to be used independently or simultaneously
with the super-beam. In the framework of R\&D for Super-Beam and Neutrino
Factory, the HARP experiment is studying hadron production at low
energies on various targets.}

\section{Introduction}
Traditionally, accelerator-based neutrino physics in Europe has been
centred around CERN. Also for the future, this laboratory is expected
to play a leading role in neutrino physics, despite the heavy 
commitment in the construction of LHC.\par
Due to the small atmospheric neutrino mass difference squared, the
attention of the neutrino community is presently shifted towards 
long-baseline project. The CERN neutrino beam to Gran Sasso is 
higher in energy with respect to the Japanese \cite{k2k} or American
\cite{numi} projects, and will have the unique feature of being well
above
threshold for $\tau$ lepton production, in order to confirm the 
$\nu_\mu\to\nu_\tau$ nature of the atmospheric oscillations.\par
The OPERA detector is a dedicated experiment to search for $\tau$
appearance using topological informations, while ICARUS is a more
versatile apparatus, that could reach similar $\tau$ identification
capabilities, provided a sufficient number of modules is built.\par
On the longer term, the European projects are focused towards the
possibility of building a super-beam based on a Superconducting
Proton Linac (SPL), that also would be the first building block of
a CERN-based neutrino factory. To estimate the flux of neutrinos
produced in a super-beam, or the number of muons accelerated in a 
neutrino factory, a dedicated experiment (HARP) is presently under 
way. The Harp detector, built in a very short time, has taken data
in summer 2001 and will have a second run in summer 2002 at the
CERN PS. The goal is to measure total and differential cross 
sections for proton interactions with a large variety of targets,
with energy in the range 3-15 GeV.

\section{The Neutrino beam from CERN to Gran Sasso}
As stated in the introduction, the CERN beam to Gran Sasso
has the unique characteristic of being at a sufficiently high 
energy to produce a detectable number of $\tau$ leptons in
charged current $\nu_\tau$ interactions. This is a natural 
consequence of using a high-energy proton driver (400 GeV from 
the CERN SPS), having as a natural target the Gran Sasso lab near
Rome (distant 732 km from CERN) and of the practical difficulties 
of building a near station, that makes hard to conceive 
disappearance measurements.\par
Civil engineering construction started on Oct 12, 2000; they will
consist of 3 km of new tunnels and caverns: 700 m for the proton
line to the target, 1 km of decay tube, plus various connections.
Overall it will mean 45000 $m^3$ of rock removal, and 12000 $m^3$
of concrete added to reinforce the walls. The main properties
of the CNGS beam are listed in table \ref{tab:beamprop}.\par
\begin{table}[tbh]
\begin{center}
\begin{tabular}{|c|c|}\hline
Baseline& 732 km\\
Maximum depth& 11.4 km\\
POT/extraction& 2.4$\times 10^{13}$\\
POT/year& 4.5$\times 10^{19}$\\
Flux at Gran Sasso&3.5$\times 10^{12}/y/100 m^2$\\
$\nu$ interactions&2500/kt/y\\
$<E_\nu>$&17 GeV\\
$\% \nu_\mu$&2.0\\
$\% (\nu_e+\bar{\nu}_e)$&0.8\\ \hline
\end{tabular}
\label{tab:beamprop}
\caption{Main properties of the CNGS beam}
\end{center}
\end{table}

The beam target will be 2 meters long, made of 10 cm long graphite
(pure carbon) rods, interspaced to minimise absorption of pions 
with small transverse momentum with respect to the beam direction. 
Due to the strong thermical shocks, the whole system is cooled 
with helium. Secondaries produced in interactions with the target
are focalised by a horn/reflector system. The horn will be 6.5 m
long, with a diameter of 70 cm, and able to sustain a current of
150 kA for 1 ms, with a tolerance rate of 95\% survival probability
after $5\times 10^7$ pulses. A horn prototype has already been 
built and tested, and survived 1.5 million pulses.\par
The beam transverse profile will be checked using two muon
detectors at the end of the decay tunnel; the beam width at the
far location is more than one kilometre, and approximately flat for
a radius of about 500 m, much larger than the expected width of the
detectors. Systematic uncertainties on the beam can arise from 
misalignment of the various components. As a reference, a 3\%
variation in the number of $\nu_\mu$ in Gran Sasso could result from:
\begin{itemize}
\item the horn being off-axis by 6 mm
\item the reflector being off-axis by 30 mm
\item the beam being off-axis by 1 mm on the target
\item an overall misalignment of 0.5 mrad
\end{itemize}
Beam alignment can be further checked in the Gran Sasso site, 
looking for the muons produced by neutrino interactions in the
upstream rock. However, we have to bear in mind that flux systematics
is not that relevant when doing an appearance experiment with few
candidate events.\par
Possible upgrade schemes for the PS/SPS complex, in order to
increase the neutrino flux, have been extensively studied 
\cite{spsupgr}. It was found out that most of the components are 
at their limits. A gain of a factor 1.5 seems feasible; however,
a factor 3 seems the absolute maximum, even if large investments
are made.\par
The Gran Sasso laboratory from INFN (Laboratori Nazionali del Gran
Sasso, LNGS) is operational since 1987, and offers an easy access 
from a motorway tunnel, and 1500 meters of rock overburden. The
completion of the first generation of experiments (e.g. MACRO) 
leaves space in two large underground hall (B and C), oriented
towards CERN, to perform long-baseline neutrino experiments.
The two experiments foreseen for the long-baseline project are
OPERA \cite{opera}, already fully approved as CNGS1, a dedicated 
experiment to look for $\tau$ kink in a hybrid emulsion system, 
and ICARUS \cite{icarus}, of which the first module has already 
been built and successfully tested, a general-purpose detector based
on a liquid Argon TPC.
\section{OPERA}
The requirement to have a large mass and a space resolution of the
order of 1 $\mu$ m has lead to the choice of hybrid emulsions as
basic detector component. A basic cell
is composed of 1 mm lead planes, separated by an emulsion layer made
of two 50 mm emulsion sheets separated by 100 $\mu$m plastic base.
56 such lead-emulsion cells constitute a brick, 3000 bricks, together
with an extruded scintillator plane for quasi-online identification
of interesting interactions are a module, 24 modules, completed with
a muon spectrometer, form a supermodule. The experiment will be
made of 3 such supermodules, for a total mass of about 2 ktons.
The emulsion surface will be 176000 $m^2$; as a comparison, that of
CHORUS was about 500 $m^2$. Given the very large amount of emulsion
surface, finding few $\tau$ events, for an average $\tau$ decay 
length of half a millimetre, is an extremely challenging task.
The first step will be the use of the scintillator tracker plane
behind the bricks to identify where neutrino interactions take place.
These bricks are removed from the wall, with a rate of about 40 bricks 
per day. They are exposed for a short period to cosmic rays in a
shallower location, always in Gran Sasso, to provide a reference 
for the alignment, and then the emulsions are developed.
Then, a search for a neutrino vertex and a decay candidate is
performed quasi-online by various scanning stations. The goal is to
understand if the brick extracted is sufficient for $\tau$ search, or
additional nearby bricks have to be extracted. Finally, about 500
squared meters of emulsions (a factor 100 more than at CHORUS) are
sent to the scanning labs.\par 
The scanning speed is a fundamental factor to ensure the success of 
the experiment. A factor 20 improvement with respect to the present
speed is expected from the use of the S-UTS technology
presently under development in the Japanese laboratory
of Nagoya University. This system is based on a fast CCD camera
(3000 frames per second) connected to microscopes, with continuous 
movement on the x-y plane. The z position is determined by the 
focusing plane of the microscopes, controlled by pizo-attenuators.
The scanning speed envisaged is 20 sqcm per hour, and the amount of
data reduction from pixel to track of the order of $10^5$.\par
The strategy for $\tau$ search is based first on finding a vertex, i.e.
a common starting point for tracks with $\tan \theta<0.4$, that are
followed back until not any more found in two successive layers.
Including geometrical losses, a vertex finding efficiency of 80\%
is expected. Then the search is split in two, depending on weather
the $\tau$ decays in the same lead layer (short decays) or in the 
next. In the first case, the requirement is to have the momentum of 
the primary track above 1 GeV, and an impact parameter larger than
a value, dependent on the z position, between 5 and 20 $\mu$m.
In case of long decays, the requirement is to have a kink angle
of 20 mrad with respect to the initial track (the kink angle 
resolution is about 3 mrad). The signal efficiency for several
event categories is shown in table \ref{tab:operaeff}, and the
number of expected backgrounds in table \ref{tab:operabg}.
\begin{table}[tbh]
\begin{center}
\begin{tabular}{|c|c|c|c|c|}\hline
&DIS long& QE long& DIS short& Overall\\ \hline
$\tau\to e$&2.7&2.3&1.3&3.4\\
$\tau\to\mu$&2.4&2.5&0.7&2.8\\
$\tau\to h$&2.8&3.5&-&2.9\\
Total&8.0&8.3&1.3&9.1\\ \hline
\end{tabular}
\label{tab:operaeff}
\caption{$\tau$ identification efficiency for OPERA in various
decay channels, for Deep Inelastic Scattering (DIS) and 
Quasi-Elastic (QE) events, in the long and short configuration}
\end{center}
\end{table}

\begin{table}[tbh]
\begin{center}
\begin{tabular}{|c|c|c|c|c|}\hline
&$\tau\to e$&$\tau\to\mu$&$\tau\to h$&Total\\ \hline
Long decays&0.15&0.29&0.24&0.67\\ \hline
Short decays&0.03&0.04&-&0.07\\ \hline
\end{tabular}
\label{tab:operabg}
\caption{Background for $\tau$ identification in number of events for
long and short decays.}
\end{center}
\end{table}
In absence of signal, OPERA can exclude at 90\% C.L. 
$\nu_\mu\to\nu_\tau$ oscillations for $\Delta m^2<1.2\times 10^{-3}
eV^2$ at full mixing, and $\sin^2 2\theta<5.7\times 10^{-3}$ at large
$\Delta m^2$. The number of expected events for signal and background
for various values of $\Delta m^2$ and maximal mixing are shown in
table \ref{tab:operaevts}
\begin{table}[tbh]
\begin{center}
\begin{tabular}{|c|c|c|c|c|}\hline
$\Delta m^2$&$1.2\times 10^{-3}$&$2.4\times 10^{-3}$&$5.4\times 10^{-3}$
&BG\\
&2.7&10.8&53.5&0.75\\ \hline
\end{tabular}
\label{tab:operaevts}
\caption{Number of expected signal events (with efficiency included)
and of background, after 5 years of CNGS operation for the OPERA 
detector}
\end{center}
\end{table}

\section{ICARUS}
The ICARUS detector is composed of a large liquid Argon TPC, that 
combines big mass and a granularity of the order of 1 mm, being
a veritable electronic bubble chamber. Its resolution and particle
identification capabilities allow this kind of detector to contribute 
to a vast and versatile physics program, that includes study of
atmospheric neutrinos, solar neutrinos, nucleon decay and of course
neutrinos from the long baseline beam.\par
Drifting electrons for distances of the order of 1 meter, needed to
achieve a sufficient detector mass, requires an extreme purity of
the liquid Argon: the tolerance is a concentration of
electronegative impurities at the level of 0.1 parts per billion,
that was finally achieved after 10 years of R\&D.\par
A 600 ton module has been built during the last years in Pavia, and
successfully tested this summer with cosmic ray data. Many long
(~20 meters) muon events were observed, as well as other 
interesting events such as
electromagnetic showers, stopping muons and strange particles.\par
This 600 ton detector will be transported to Gran Sasso at the beginning
of 2003. The ICARUS technology is very well-suited for $\tau$ search,
using a kinematic approach, complementary to the topological one of
OPERA. However, 600 tons are not sufficient to produce enough events
when the tight cuts necessary for large background suppression are
applied. The ICARUS collaboration has recently proposed 
\cite{icarusprop} an upgrade to 5 modules, for a total fiducial mass of 2.35 kton. Golden $\tau\to e$ events can be identified cutting on 
fiducial volume, electron momentum, visible energy, electron $P_t$
and $Q_t$ and missing momentum. The best sensitivity is obtained
combining several variables in a common likelihood, and efficiencies 
and backgrounds for a 3 kton ICARUS (table \ref{tab:icatau}) is 
similar to that of OPERA.
\begin{table}[tbh]
\begin{center}
\begin{tabular}{|c|c|c|c|c|c|}\hline
$\tau$ decay mode&1.6&2.5&3.0&4.0&BG\\ \hline
$\tau\to e$&3.7&9&13&23&0.7\\
$\tau\to\rho$ DIS&0.6&1.5&2.2&3.9&$<0.1$\\
$\tau\to\rho$ QE&0.6&1.4&2.0&3.6&$<0.1$\\
Total&4.9&11.9&17.2&30.5&0.7\\ \hline
\end{tabular}
\label{tab:icatau}
\caption{Signal events for different values of $\Delta m^2$ (in units
of $10^{-3} eV^2$) and
expected number of background events after 5 years of data taking
for a 3 kton ICARUS detector.}
\end{center}
\end{table}
Due to its good electron identification, ICARUS will also be able to 
perform a search for $\nu_\mu\to\nu_e$ oscillations, improving by some
factors the present limit on $\sin^2 2\theta_{13}$.

\section{The CERN-Frejus Super-Beam}
In the next 15-20 years the main commitment of CERN will be the 
construction and operation of LHC and its detectors. However, the 
European neutrino community is very active in trying to shape a
next-generation neutrino program, in the context of a future CERN-based
neutrino factory. The first brick of such a machine would be a low
energy (2.2 GeV) high power (4 MW) proton linac\cite{spl}.\par
Independently of the neutrino factory, this machine would be a 
considerable improvement to the CERN accelerator complex, since it
will increase PS beam intensity, triple the brilliance of the LHC beam,
supply the present ISOLDE facility with 5 times more current (or up to
100 times for Super-ISOLDE), and allow the construction of a low-energy
superbeam.\par
Focusing the pions in the energy range of their maximal production,
the resulting mean neutrino energy is about 250 MeV, an ideal energy for
a water Cerenkhov detector. At this energy, the maximum of the 
oscillation occurs for a baseline of about 130 Km, the distance between
CERN and the Frejus tunnel. Due to the civil engineering
work foreseen to build a second motorway gallery, it is conceivable
to propose building there a large Cerenkhov detector, to be used for
proton decay, supernova and beam neutrinos, like that proposed by the
UNO collaboration \cite{uno}. Considering 400 kton of fiducial mass,
a $\pi^0$/e separation with 0.1 \%
confusion and efficiency around 70\%, the 90\% C.L. sensitivity to 
$\sin^2 2\theta_{13}$ would be in the range of $5\times 10^{-4}$ 
after 5 years for the current values of $\Delta m^2$. A ten times 
smaller exposure would already allow measuring $\sin^2 \theta_{23}$ 
at 1\%
level and $\Delta m^2_{23}$ with a precision of $10^{-4}$. 
For CP violation studies, even in the case of a favourable choice of
parameters, 400 kton are needed, for 2 years of running with neutrinos
and 10 years with antineutrinos.
\section{The beta-beam}
A recent interesting possibility\cite{zuc} is producing a neutrino
beam from the decay of radioactive ions collected in a storage ring.
Antineutrinos would be for instance produced by the decay
$^6He^{++}\to^6Li^{+++} \bar{\nu}_e e^-$, and neutrinos from similar
decays of $^{18}Ne$. The advantage of this approach with respect to
the ``traditional'' Neutrino factory is that despite the lower 
intensities of the stored beam, the much larger quality factor
$\Gamma/E_0$ for ions produces a very collimated neutrino beam,
leading to fluxes in the far detector similar to those of a super-beam,
for comparable energies. The produced beam will 
be extremely clean and well-known, and of only the electron neutrino
flavour. Moreover, it is possible to think of an experiment combining
the beta-beam and the super-beam, that would produce an almost pure
$\nu_\mu$ beam, separated from the $\nu_e$ beta-beam
using timing information. Such a system would have the two-beam feature
of the neutrino factory, but it will not need a magnetic detector to
distinguish oscillated and beam events. Therefore, larger detector
with lower threshold can be used (like a water Cerenkov), putting
them at shorter distance to minimise matter effect. A high-intensity
combination of super-beam and beta-beam could search for T-violation in
an extremely convincing way.
\section{Other experiments}
The HARP detector, presently running at CERN, is not strictly speaking
a neutrino experiment, but it will provide measurement of hadron
production at small and large angles for a variety of target material
of interest for neutrino beams. In particular large angle pion
production is important in the design of the focusing system of a
Neutrino Factory and a Super-Beam. It will also use cryogenic targets
made of liquefied gases for hadronic production in the atmosphere 
in the region of interest of atmospheric neutrinos, as well as the
measurement of replicas of existing neutrino targets (K2K, MiniBOONE).
Several million of events have been collected in summer 2001, and 
a similar statistics will be collected this year. Since several
sources of data-taking inefficiency have been removed, the number of
useful events is expected to increase by a factor of approximately 5.
\par
The possibility of using the existing AD ring as a muon accumulator
for a ``baby'' neutrino factory to measure low-energy neutrino cross 
section was studied \cite{nucross}. The interesting result was that
these muons are already accumulated in the present antiproton running
mode, and thus a neutrino beam is already existing there. Unfortunately,
quantitative estimate showed that obtaining a sufficient number of 
stored muons in parasitic mode is incompatible with the present
setup of the running AD experiments.
\section{Conclusions}
In the next years, CERN will focus on building and operating LHC.
However, a large community is eager to maintain an active neutrino 
physics program. The CNGS long-baseline beam will start operation in
spring 2005; the OPERA detector is already fully approved and under
construction; a first 600 ton ICARUS module has already bee built and
tested, and a proposal for a factor 5 mass increase is under approval.
For the future, HARP is studying hadron production in the region of
interest of present and future neutrino beams, neutrino factories
and atmospheric neutrinos; there is an active group studying the
possibility of a second-generation super-beam (possibly in conjunction
with a beta-beam), and trying to make the first step towards a European
neutrino factory for the post-LHC period.

\end{document}